\documentclass[a4paper,11pt]{article}
\usepackage{jheppub} 
\usepackage{tikz}
\usepackage{simplewick}
\usepackage{ulem}
\newcommand{\be}{\begin{equation}}
\newcommand{\ee}{\end{equation}}
\newcommand{\bea}{\begin{eqnarray}}
\newcommand{\eea}{\end{eqnarray}}

\newcommand{\ba}{\begin{aligned}}
\newcommand{\ea}{\end{aligned}}
\newcommand{\p}{\partial}

\title{ Correlation Function Of Thin-Shell Operators}

\author[a,b,c]{Bin Chen}
\author[a]{Yuefeng Liu}
\author[b]{and Boyang Yu}
\affiliation[a]{School of Physics, Peking University, No.5 Yiheyuan Rd, Beijing 100871, P.R. China}
\affiliation[b]{Center for High Energy Physics, Peking University, No.5 Yiheyuan Rd, Beijing 100871, P.
R. China}
\affiliation[c]{Collaborative Innovation Center of Quantum Matter, No.5 Yiheyuan Rd, Beijing 100871,
P. R. China}

\emailAdd{bchen01@pku.edu.cn}
\emailAdd{yfliu0905@pku.edu.cn}
\emailAdd{yuby21@pku.edu.cn}

\abstract{In this study, we explore the correlation functions of thin-shell operators, represented semiclassically by a homogeneous, thin interface of dust particles. Employing the monodromy method, we successfully compute the {contribution from the Virasoro vacuum block and present the monodromy equation in a  closed form without assuming  the probe limit.} Although an analytical solution to the monodromy equation remains difficult, we demonstrate {that it is perturbatively solvable} within specific limits, including the probe limit, the heavy-shell limit, and the early-time limit. Moreover, we compare our results with gravitational calculations and find precise agreement. We {strengthen our findings by proving that the thermal correlation functions in gravity, after an inverse Laplace transformation, satisfy the field theory's monodromy equation.} Additionally, we identify an infinite series of unphysical solutions to the monodromy equation and discuss their potential geometrical duals.

}

\begin{document}
\maketitle
\flushbottom

\section{Introduction}

In the simplified version, AdS/CFT correspondence states that certain bulk quantum gravity in an Anti-de Sitter (AdS) spacetime could be equivalent to a conformal field theory(CFT) at the AdS boundary  \cite{Maldacena:1997re}. It identifies the partition function  of bulk quantum gravity $Z_{\text{grav}}[\mathcal B]$,  where $\mathcal B$ denotes the spacetime manifold, with the one of holographic CFT $Z_{\text{CFT}}[M]$ defined on the boundary $\mathcal M=\partial\mathcal B$  \cite{Witten:1998qj},
\be
Z_{\text{CFT}}[\mathcal M]=Z_{\text{grav}}[\mathcal B].
\ee
At the semiclassical level, the gravitational path integral is approximated by a sum of the on-shell actions of saddle-point geometries. Usually the summation is dominated by one of  the saddle points $\mathcal B_0$ such that  \cite{Gibbons:1976ue,Hawking:1982dh}
\be
Z_{\text{grav}}[\mathcal B]\approx e^{-I[\mathcal B_0]}.
\ee
The correspondence allows us to compute the correlation function holographically. If the operators in CFT are light, we may include a source in the bulk action and compute the correlation functions in the probe limit  \cite{Witten:1998qj}. However, when the operators are heavy enough, we have to start from the bulk action  
\be
I=I_{gravity}+I_{matter},\nonumber
\ee 
to find the saddle-point geometries, {taking into account of} the backreaction of matter to the background spacetime. 

The semiclassical picture of the AdS/CFT correspondence is particularly clear in the case of AdS$_3$/CFT$_2$. The semi-classical AdS$_3$/CFT$_2$ correspondence states that semiclassical AdS$_3$ gravity with the Brown-Henneaux asymptotic boundary condition \cite{Brown:1986nw} is dual to holographic two-dimensional CFT with a large central charge \cite{Strominger:1997eq,Hartman:2014oaa}
\be
c={\frac{3\ell}{2G}},
\ee
where $G$ is the three-dimensional gravitational coupling constant and $\ell$ is the AdS radius. The AdS$_3$ gravity is more tractable than its higher-dimension cousins  \cite{Achucarro:1986uwr,Witten:1988hc,Witten:2007kt}, and it allows us to study the backreaction and 1-loop quantum correction in a controllable way  \cite{Krasnov:2000zq,Yin:2007gv,Giombi:2008vd}. The semiclassical AdS$_3$/CFT$_2$ correspondence has played important roles in the study of entanglement entropy  \cite{Ryu:2006bv,Ryu:2006ef,Headrick:2010zt}. In this case, one often has to deal with the backreacted geometries and even the 1-loop quantum corrections \cite{Hartman:2013mia,Faulkner:2013yia,Barrella:2013wja,Chen:2013kpa,Chen:2013dxa,Chen:2014kja,Chen:2014unl,Headrick:2015gba,Chen:2015uia,Chen:2015kua,Chen:2015uga,Zhang:2015hoa,Li:2016pwu,Chen:2016dfb,Chen:2016kyz,Chen:2016lbu,Chen:2016uvu,Belin:2017nze}. 



Studying correlation function in high energy states has drawn much attention due to its relation {with} the Eigenstate Thermalization Hypothesis (ETH), which is a { significant} concept in quantum mechanics and statistical physics {and offers} a theoretical explanation for the thermalization process in isolated quantum systems \cite{Srednicki:1994mfb,Deutsch:1991msp,Rigol:2007juv,DAlessio:2015qtq}.
It posits that the states with sufficiently small energy fluctuations are nearly thermal and  the observables computed in such states are close to those of micro-canonical ensembles. The above statements can be simply formulated as 
\be\label{ETH0}
\langle \mathcal O_{obs}\rangle_E\approx\langle \mathcal O_{obs}\rangle_{\beta_E}.
\ee
 In AdS$_3$/CFT$_2$, by setting  $\mathcal O_{obs}=O_LO_L$ and requiring $h_L/c\ll1,h_H/c\gg 1$ in the holographic CFT, we are led to the heavy-heavy-light-light limit, which is the case most relevant for probing ETH \cite{Hartman:2013mia,Asplund:2014coa,Fitzpatrick:2015zha,Balasubramanian:2017fan}. The heavy state corresponds to the black hole in the bulk. For various aspects on ETH in AdS/CFT, see  \cite{Lashkari:2016vgj,Dymarsky:2016ntg,Lin:2016dxa,He:2017txy,He:2017vyf,Lashkari:2017hwq}. {The relation \eqref{ETH0} seems to raise the issue of forbidden singularity: the correlator on the left-hand side (LHS) of \eqref{ETH0} should  possess the OPE singularity only; However, the thermal correlator on the right-hand side (RHS) is singular at both the OPE singularity and its thermal images, and the extra singularities from these thermal images are forbidden according to \eqref{ETH0}.
Since the relation \eqref{ETH0} holds only in the probe limit and to the leading order in the large $c$ expansion, the issue of forbidden singularity is expected to be resolved by summing over all possible corrections including the finite-$c$ effect and the finite-probe effect. In particular, it has been argued that studying the finite-probe effect would be an important step towards the finite-$c$ effect \cite{Faulkner:2017hll}.}

 {The thin-shell operator is a special kind of non-local operator, and has played important roles in the AdS/CFT correspondence.}  In a holographic CFT$_d$ defined on $\mathbb R_t\times S^{d-1}$, a thin-shell operator consists of the product of $O(N^2)$ local primary operators smearing homogeneously on the sphere
\be
\mathcal V=\prod_{i=1}^n\psi(t,\Omega_i).
\ee
The number of local primaries scales with the central charge $n\sim N^2\sim c$ so that it admits a semiclassical bulk description: a  spherical thin interface of {falling dust particles which is heavy enough to  backreact to the geometry. The thin-shell operator has been considered in 2D holographic CFT \cite{Anous:2016kss} to create a collapsed state, corresponding to the spherical collapse of a shell of matter in AdS$_3$. It has been shown that  the correlation function of two local operators in such state agrees perfectly with the gravitational calculations in AdS-Vaidya geometry. Recently, such operators were used to construct black hole microstates \cite{Balasubramanian:2022gmo,Balasubramanian:2022lnw,Climent:2024trz}. Besides, due to its convenience in constructing the wormhole solutions in higher dimensions, the shell operators can be modeled by the ETH ensemble \cite{deBoer:2023vsm}. In  \cite{Sasieta:2022ksu,Antonini:2023hdh}, the authors used the thin-shell operators  to study semiclassical bulk wormhole solutions.

In this paper, we study the correlation function of the thin-shell operators in the {holographic CFT$_2$. In particular, we compute the contribution from the  Virasoro vacuum block by using the monodromy method.} In this case, we find that the monodromy equation can be obtained for finite $\hat e_\psi/e_H$ without {the need to consider the probe limit, where the parameters $e_H$ and $\hat{e}_\psi$ are defined by $e_H\equiv 6h_H/c$ and $\hat{e}_\psi\equiv 6 n h_\psi/c$ with $h_H$, $h_\psi$ being the conformal dimensions of operators and  $c$ being the central charge}.  Although the equation cannot be solved exactly, we can study it in various limits, such as the probe limit $\hat e_\psi/e_H\ll 1 $, the heavy-shell limit $\hat{e}_\psi/e_H\gg1$ or the early-time limit. We will discuss these three cases and solve the monodromy equation perturbatively.  By moving away from the probe limit, we could see how the ETH \eqref{ETH0} gets modified.

The paper is organized as follows: In section 2, we use the monodromy method to study the Virasoro vacuum block for the correlator of the thin-shell operators and obtain the monodromy equation. In section 3, we review the computation of thermal correlation function of the thin-shell {operators} in gravity. In section 4, we solve both the monodromy equation and gravitational saddle-point equation in  {three different limits and show that the perturbative solutions on both sides are in agreement.}  Then we give a general non-perturbative proof to show that the microcanonical correlation function obtained by using the monodromy method is indeed related to the thermal correlation function obtained in gravity via an inverse Laplace transformation.  {Furthermore  we analyze the additional unphysical solutions to the monodromy equation and give them semiclassical geometrical descriptions. In section 5, we end with conclusions and some discussions.}

\section{{Vacuum Virasoro block}}

\begin{figure}
    \centering
    \includegraphics[width=0.7\textwidth]{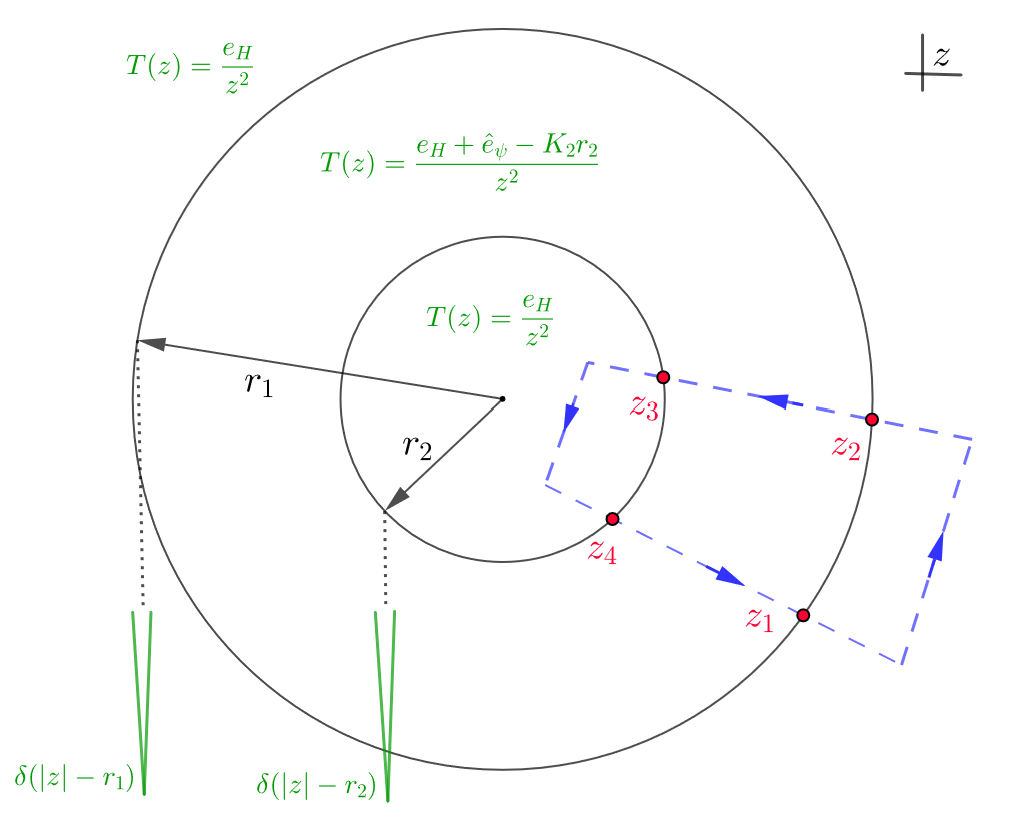}
    \caption{ {The setup for the differential equation \eqref{dif-eq} and the monodromy condition \eqref{contourf} in computing the correlator \eqref{shellfield} of the thin-shell operators. The dashed closed loop in blue denotes the monodromy contour, two $\delta$-functions in green appear in $T_{\psi}$ in \eqref{potentialT}. }}
    \label{fig:monodromy}
\end{figure}
    

In the holographic CFT$_2$, {which has a large central charge $c\gg1$ and sparse light spectrum,} defined on a cylinder, we aim to study the correlation function of the thin-shell operators in a heavy state, 
\be \label{shellfield}
G(t_1-t_2):=\langle h_H|\mathcal{V}(t_1)\mathcal{V}(t_2)|h_H\rangle,\quad t_1>t_2,
\ee
where the thin-shell operator $\mathcal V$ is given by the product  of infinite identical local operators distributing evenly on the constant time slice, i.e.
\be
\mathcal V(t_i)=\prod_{k=1}^n\psi(t_k,e^{2\pi i(k-1)/n}),\quad n\to\infty.
\ee
Performing an exponential map to transform the cylinder to the plane, the correlator becomes
\be\ba\label{corre}
G(t_1-t_2)&=\prod_{i=1}^n(z_1^{(i)}z_2^{(i)})^{2h_\psi}\langle O_H(\infty)\prod_{i=1}^n\psi(z_1^{(i)})  \psi(z_2^{(i)})O_H(0)\rangle
\\&=e^{2nh_\psi(t_1+t_2)}\langle O_H(\infty)\prod_{i=1}^n\psi(z_1^{(i)})  \psi(z_2^{(i)})O_H(0)\rangle,
\ea
\ee
with
\be
z_1^{(k)}=r_1e^{2\pi i(k-1)/n},\quad z_2^{(k)}=r_2e^{2\pi i(k-1)/n},\quad r_1>r_2
\ee
and $r_i=e^{t_i}$.

 {As shown in  \cite{Hartman:2013mia}, the correlation functions in the holomorphic CFT$_2$ are dominated by the vacuum Virasoro conformal block, and can be  computed using the monodromy method, which was summarized in  \cite{Zamolodchikov:1987avt,Harlow:2011ny,Hartman:2013mia} and further developed in  \cite{Anous:2016kss}}.  To proceed, we consider the following differential equation
\be\label{dif-eq}
V''(z)+T(z)V(z)=0,
\ee
where the associated stress tensor is given by
\be\label{stress}
T=\frac{e_H}{z^2}-\frac{c_H}{z}+\sum_{j=1}^2\sum_{i=1}^n\frac{e_\psi}{(z-z_j^{(i)})^2}-\frac{c_{\psi,i}}{z-z_j^{(i)}}\equiv T_H+T_\psi,
\ee
with $e_H=6h_H/c,e_\psi=6h_\psi/c$ and $c_H,c_{\psi,i}$ being accessory parameters which will be fixed by the boundary condition and the monodromy condition.
In the continuous limit, the infinite summation in \eqref{stress} can be written as an integral
\be
T_\psi=\sum_{j=1}^2\int_0^{2\pi}\frac{d\theta}{2\pi}\big[\frac{\hat e_\psi}{(z-r_je^{i\theta})^2}-\frac{c_{\psi,j}}{z-r_je^{i\theta}}\big]
\ee
with $\hat e_\psi=ne_\psi$.
The angular dependence of $c_{\psi,j}$ is fixed by requiring the residue of $T_\psi$ to be independent of $\theta$ in cylinder coordinates due to rotational symmetry, so that we have $c_{\psi,j}=K_je^{-i\theta}$. In the large $c$ limit, 
the correlation function on plane can be approximated by
\be\label{largec}
\langle O_H(\infty)\prod_{i=1}^n\psi(z_1^{(i)}) \prod \psi_{i=1}^n(z_2^{(i)})O_H(0)\rangle\approx e^{-\frac{c}{3}f}
\ee
with $f$ satisfying
\be\label{derivative}
\frac{\p f}{\p r_i}=K_i~~\Leftrightarrow~~ \frac{\p f}{\p t_i}=K_ir_i.
\ee
Using the following  integrals
\be\ba
&\int^{2\pi}_0\frac{d\theta}{2\pi}\frac{1}{(z-re^{i\theta})^2}=\frac{d}{dr}\int^{2\pi}_0\frac{d\theta}{2\pi}\frac{e^{-i\theta}}{z-re^{i\theta}},\\
&\int^{2\pi}_0\frac{d\theta}{2\pi}\frac{e^{-i\theta}}{z-re^{i\theta}}=\frac{r\Theta(|z|-r)}{z^2},
\ea
\ee
we obtain the expression for $T_\psi$
\be \label{potentialT}
T_\psi=\frac{(\hat{e}_\psi-K_1r_1)\Theta(|z|-r_1)+(\hat{e}_\psi-K_2r_2)\Theta(|z|-r_2)}{z^2}-\frac{\hat{e}_\psi[r_1\delta(|z|-r_1)+r_2\delta(|z|-r_2)]}{z^2}.
\ee
One peculiar property of the stress tensor we have obtained is that in the continuous limit, its singular behavior takes the form of  {two delta functions on the ring $|z|=r_i, i=1,2$, as shown in Fig. \ref{fig:monodromy}, together with a double pole at $z=0$.} This prevents us from solving a differential equation with infinite singular points, but enables us to solve the ODE \eqref{dif-eq} exactly.
Before solving \eqref{dif-eq}, we need to reduce the number of parameters by imposing proper boundary conditions.
Since we have inserted the operator $O_H$ at the infinity to create the heavy state,
the asymptotic behavior of the stress tensor at $|z|\to\infty$ is required to be $T\sim\frac{e_H}{z^2}$ which implies
\be\label{constraint}
c_H=0,\quad 2\hat{e}_\psi-K_1r_1-K_2r_2=0.
\ee
This is consistent with the time-translation symmetry. To see this, we can combine \eqref{corre} with \eqref{largec} to get
\be\label{f-G}
f\approx-\frac{3}{c}\log\big(e^{-2nh_\psi(t_1+t_2)}G(t_1-t_2)\big),
\ee
so {its derivatives satisfy}
\be
K_1r_1+K_2r_2=\frac{\p f}{\p t_1}+\frac{\p f}{\p t_2}=\frac{12nh_\psi}{c}=2\hat{e}_\psi
\ee
which is just the second equation of  \eqref{constraint}. In conclusion, the stress tensor is given by
\be
T(z)=\left\{
\begin{array}{lc}
  \frac{e_H}{z^2},   & ~~|z|>r_1, \\
     \frac{e_H+\hat{e}_\psi-K_2r_2}{z^2}- \frac{\hat{e}_\psi }{z^2} (r_1\delta(|z|-r_1)+r_2\delta(|z|-r_2),  & ~~r_2\leq|z|\leq r_1,\\
  \frac{e_H}{z^2},  &~~ |z|<r_2.
\end{array}
\right.
\ee

We are now ready to study the monodromy condition.
The solution to the differential equation \eqref{dif-eq}
is  given by
\be\label{sol}
V(z)=\left\{
\begin{array}{cc}
   J_1V_1,  &~~ 0<|z|<r_2, \\
    V_2, &~~ r_2<|z|<r_1,\\
    J_2V_3,&~~ |z|>r_1,
\end{array}
\right.
\ee
{with $V_i$ being two branches of linearly independent solutions in the corresponding regions. The explicit expressions of $V_i$ are}
\be
V_i=(z^{1/2-\rho_i},z^{1/2+\rho_i})
\ee
where
\be\label{rho12}
\rho_1=\rho_3=\frac{\sqrt{1-4e_H}}{2},\quad \rho_2=\frac{\sqrt{1-4(e_H+\hat{e}_\psi-K_2r_2)}}{2}.
\ee
   {The matrices $J_1$ and $J_2$ in \eqref{sol} are  determined by} the junction conditions across the shells. More precisely,
the  solution  \eqref{sol} is required to be continuous across the two shells
\be
J_1V_1-V_2\big|_{|z|=r_2}=0,\quad J_2V_3-V_2\big|_{|z|=r_1}=0,
\ee
but   {the firs derivative of $V(z)$ jumps at $z=r_1e^{i\theta}$} according to
\be\ba
&V'(z_1^{(+)})-V'(z_1^{(-)})=\int^{z=z_1^{(+)}}_{z=z_1^{(-)}} dz V''
=\frac{\hat{e}_\psi e^{-i\theta}}{r_1}V(r_1e^{i\theta}),
\ea\ee
which implies
\be
J_2V_3'-V_2'|_{z=r_1e^{i\theta}}=\frac{\hat{e}_\psi}{z}V(z)\big|_{z=r_1e^{i\theta}}.
\ee
Similarly, we have
\be
V_2'-J_1V_1'\big|_{z=r_2e^{i\theta}}=\frac{\hat{e}_\psi}{z}V(z)\big|_{z=r_2e^{i\theta}}.
\ee
With these, $J_1$ and $J_2$ are solved to be of the following forms
\be
\ba
&J_1=\frac{1}{2\rho_1}\left(
\begin{array}{cc}
z^{\rho_1-\rho_2}(\hat e_\psi+\rho_1+\rho_2)&z^{-\rho_1-\rho_2}(-\hat e_\psi+\rho_1-\rho_2)\\
z^{\rho_1+\rho_2}(\hat e_\psi+\rho_1-\rho_2)&z^{-\rho_1+\rho_2}(-\hat e_\psi+\rho_1+\rho_2)
\end{array}
\right),\quad |z|=r_2,
\\&J_2=\frac{1}{2\rho_1}\left(
\begin{array}{cc}
z^{\rho_1-\rho_2}(-\hat e_\psi+\rho_1+\rho_2)&z^{-\rho_1-\rho_2}(\hat e_\psi+\rho_1-\rho_2)\\
z^{\rho_1+\rho_2}(-\hat e_\psi+\rho_1-\rho_2)&z^{-\rho_1+\rho_2}(\hat e_\psi+\rho_1+\rho_2)
\end{array}
\right),\quad |z|=r_1.
\ea
\ee
The solution may receive a monodromy as it goes around a closed loop crossing the shells. Specifically   {with the chosen loop similar to the one in  \cite{Anous:2016kss}, see Figure \ref{fig:monodromy},} the monodromy matrix is given by
\be \label{contourf}
M=J_2(z_1)J_2^{-1}(z_2)J_1(z_3)J_1^{-1}(z_4),
\ee
where
\be
z_1=r_1e^{i\theta_1},\; z_2=r_1e^{i\theta_2},\; z_3=r_2e^{i\theta_2},\; z_4=r_2e^{i\theta_1}.
\ee
Since we are interested in   {the contribution from} the vacuum conformal block,
 the monodromy matrix is supposed to be trivial and $M=id$ gives the monodromy equation
\be\label{mono-eq}
r_2^{2\rho_2}(\rho_2+\rho_1-\hat{e}_\psi)(\rho_2-\rho_1-\hat{e}_\psi)=r_1^{2\rho_2}(\rho_2-\rho_1+\hat{e}_\psi)(\rho_1+\rho_2+\hat{e}_\psi).
\ee
This is an algebraic equation for the time derivative of the correlator. Remarkably, we have obtained the monodromy equation without turning to any limit such as the HHLL limit.

\section{{Thermal correlator in gravity}}

In this section, we review the computation of  thermal correlator of the thin-shell operator in the $AdS_3$ gravity, following  \cite{Sasieta:2022ksu}. {The thin-shell operator can be holographically described} by {a} spherical cloud of dust particles, which backreacts on the geometry classically. The dust cloud can be effectively described by a perfect fluid with the stress tensor
\be
T_{\mu\nu}\big|_{\mathcal W}=\sigma u_\mu u_\nu,
\ee
{and its total mass is}
\be \label{mass}
m=2\pi r_\infty \sigma(r_\infty)\; ,
\ee
where $u^\mu$ denotes proper velocity of the fluid, $\sigma(r)$ denotes the fluid density at radius $r$ and $r_\infty$ is the cutoff radius. The backreacted geometry is the classical solution of {the Euclideanized} action
\be
I=-\frac{1}{16\pi G}\int\sqrt{g}(R-2\Lambda)+\int_{\mathcal W}\sqrt{h} \sigma \; .
\ee
  {In this work, we set the AdS$_3$ radius to be unit $\ell=1$. The saddle-point solution related to the thermal correlator $G_\beta(t)$ contains the worldvolume $\mathcal W$ of the perfect fluid, which starts from the asymptotic boundary $r_\infty$, propagates inward until reaching the turning point $r_*$,
  then returns to the asymptotic boundary. The turning point $r_*$ \be r_*^2=r_{\pm}^2+\big( \frac{M_{\pm}-M_{\mp}}{m}-2 G m \big)^2 \ee 
 satisfies $V_{\text{eff}}(r_*)=0$ in \eqref{Veff}.}  The {worldvolume} $\mathcal W$ divides the saddle-point manifold into two pieces, see Figure \ref{partition}.
Away from $\mathcal W$, each piece is locally AdS$_3$ and is described by a BTZ metric
\be
ds^2_\pm=f_\pm(r)dt_\pm^2+\frac{dr^2}{f_\pm(r)}+r^2d\phi^2 \;,
\ee
where 
\be
f_\pm(r)=r^2-8GM_\pm
\ee
is the blackening factor with $M_\pm$ being the ADM mass of the BTZ black hole. The horizon radii $r_\pm$ and the inverse temperatures $\beta_\pm$ can be obtained via
\be\label{relation}
M_\pm=\frac{r^2_\pm}{8G}=\frac{\pi^2}{2G\beta^2_\pm} \; .
\ee 
These two black holes are glued together via the Israel junction condition along the shell $\mathcal W$, whose trajectories $(r(T),t_\pm(T))$ in terms of the proper time $T$ satisfy   
\be
\label{shell-eq}
f_\pm(r) \, \dot{t}_\pm=\sqrt{f_\pm(r)-\dot{r}^2}\, ,\quad 
\dot{r}^2+V_{\text{eff}}(r)=0 \; ,
\ee
where the effective potential $V_{\text{eff}}$ is given by
\be\label{Veff}
V_{\text{eff}}(r)=-f_{{\pm}}(r)+\left(\frac{M_{{\pm}}-M_{{\mp}}}{m}-2Gm\right)^2 \; .
\ee
Also the time elapsed by the shell in each BTZ patch were computed in  \cite{Sasieta:2022ksu} by using \eqref{shell-eq}
\be
\Delta t_\pm=2\int^{r_\infty}_{r_*}\frac{dt}{dr}dr=2\int^{r_\infty}_{r_*}\frac{dr}{f_\pm}\sqrt{\frac{f_\pm+V_{\text{eff}}}{-V_{\text{eff}}}}
=\beta_\pm\frac{\arcsin(r_\pm/r_*)}{\pi} \; .
\ee

\begin{figure}\label{manifold}
    \centering
    \includegraphics[width=0.49\textwidth]{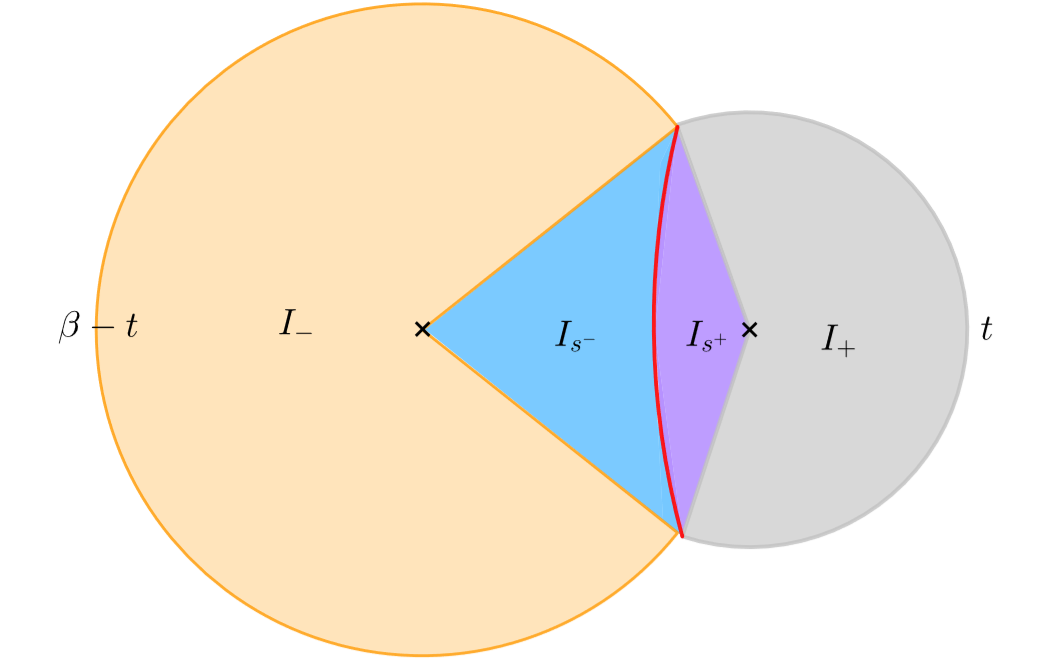}
     \includegraphics[width=0.40
    \textwidth]{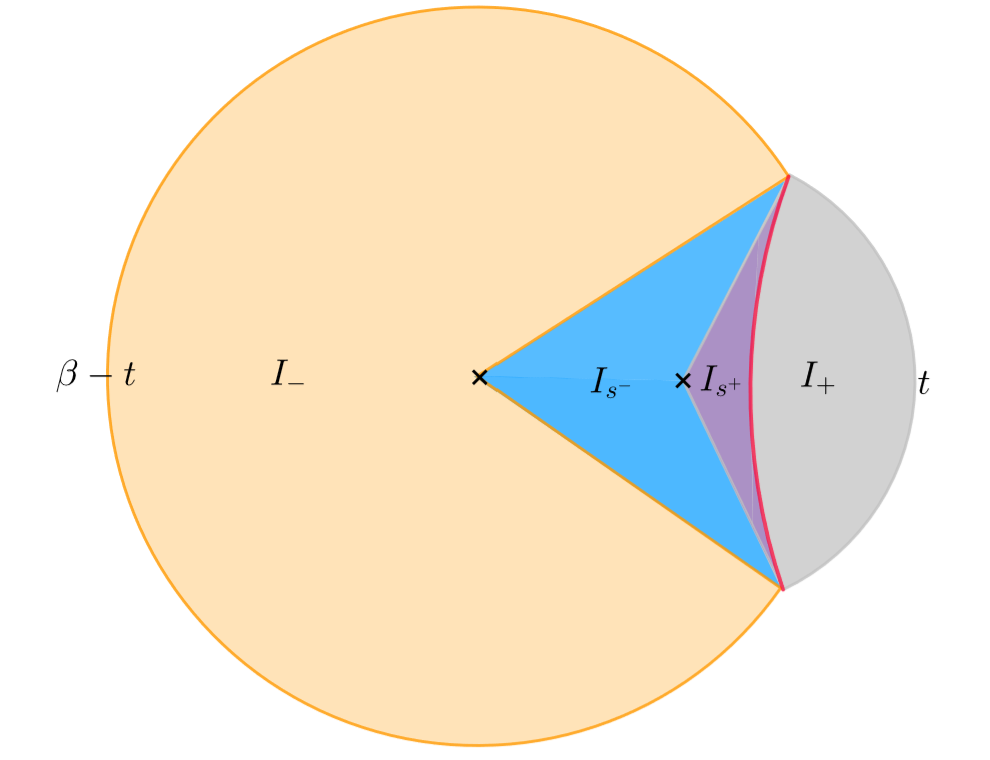}
    \caption{  {Backreacted geometries for $t\geq t_c$ (left) and $0\leq t<t_c$ (right):  the red lines denote the worldvolume $\mathcal W$, the blue regions are bounded by $\mathcal W$ as well as two radii of (orange) BTZ with mass $M_{-}$, and the purple regions are bounded by $\mathcal W$ as well as two radii of (grey) BTZ with mass $M_{+}$. The gravitational actions (purple) $I_{s^+}$ and (blue) $I_{s^-}$ ``together'' denote the intrinsic contribution of the shell.}}
    \label{partition}
\end{figure}

The backreacted manifold can be classified into two kinds\footnote{Throughout the paper, we assume $M_+>M_-$ corresponding to $t\leq\frac{\beta}{2}$. The complementary regime can be obtained by a symmetric replacement $t\to\beta-t$. }, depending on the value of $t$ as illustrated in Fig. \ref{partition}. For    $0<t<t_c$, the right patch does not include $r=r_+$ tip of the cigar geometry. As $t$ increases, $M_+-M_-$ decreases. When $t> t_c$, the right patch would include $r=r_+$ tip. Here $t_c$ is the critical time when the turning point $r_*$ coincides with the right tip $r_+$. The potential \eqref{Veff} tells us that $t=t_c$ is equivalent to $M_+-M_-=2Gm^2$, thus the condition $0<t<t_c$ is equivalent to the condition $M_+-M_->2Gm^2$, and the condition $t> t_c$ corresponds to $M_+-M-< 2Gm^2$. The input data of this setup includes $\{\beta,t,m \}$, where $\beta$ is the inverse temperature of original black hole with mass $M$, $t$ is the insertion time  of the dust particles and $m$ is the total mass of the dust particles \eqref{mass}. The backreacted geometry is determined by the data and eventually characterized by $\{ M_+, M_- \}$. The equations connecting $\{ M_+, M_- \}$ and $\{\beta,t,m \}$ are
\begin{align}
    & \beta_-=\beta-t+\Delta t_- \, , \quad \Delta t_+=t \, , \quad \quad \quad \quad 0<t<t_c \, , \label{eq1} \\
   & \beta_-=\beta-t+\Delta t_- \, ,\quad \beta_+=t+\Delta t_+ \, , \quad \quad \quad t> t_c \, . \label{eq2}
\end{align}

  {There appear two kinds  of backreacted geometries, depending on the elapsed time, as shown in Fig. \ref{partition}. Nevertheless, they can be studied in a uniform way.  Firstly the gravitational on-shell action $I$ is a summation of four terms} 
\begin{align} 
& I=I_- + I_+ + I_{s^+} - I_{s^-} \, , \quad \mbox{when}~~M_+-M_->2Gm^2 \\
& I=I_- + I_+ + I_{s^+} + I_{s^-} \, , \quad  \mbox{when}~~ M_+-M_-<2Gm^2 
\end{align}
where {the actions $ I_{s^{\pm}}$ from the shell} are respectively
\be I_{s^+}=\left| \frac{M_{+}-M_{-}}{2 G m}-m  \right| \cosh^{-1}{\big(\frac{r_\infty}{r_*} \big) }\, , \quad I_{s^-}=\left| \frac{M_{-}-M_+}{2 G m}-m  \right| \cosh^{-1}{\big(\frac{r_\infty}{r_*} \big) }. \ee 
  {In any case, the intrinsic contribution of the shell can be summarized in one formula $I_s$ (or its renormalized version)
\be
I_{s}= 2m \cosh^{-1}{\big(\frac{r_\infty}{r_*} \big) } \sim -2 m \log{r_*}+ 2m \log{2}   \, ,
\ee
and the gravitational action is simply 
\be \label{action} I=I_- + I_+ + I_{s}. \quad   \ee 
Moreover, both \eqref{eq1} and \eqref{eq2} can be rewritten as 
\be\label{grav-eq}
\frac{r_-}{r_*}=\sin\frac{r_-(\beta-t)}{2}\, , \quad  \frac{r_+}{r_*}=\sin\frac{r_+t}{2} \, .\ee
In other words, we do not need to worry about different geometries and may work directly on the action \eqref{action} and the relation \eqref{grav-eq}. This fact  would play significant role in proving ``field=gravity'' {in the context of the correlators of the thin-shell operators via AdS/CFT in the next section. }  

We list other formulas useful {in the following discussion.} The gravitational actions $I_{\pm}$ are 
\be \label{act3} I_+=t F(\beta_+)\, , \quad I_-=(\beta-t)F(\beta_-)\, , \quad F(\beta_\pm)=-M_{\pm} \, ,\ee 
where $F(\beta_\pm)$ is the free energy\footnote{  {Note that there is a typo in the free energy $F(\beta_\pm)$ in  \cite{Sasieta:2022ksu}.} }. The thermal two-point function of the thin-shell operators is captured by 
\be\label{Gbeta}
G_\beta(t)\approx e^{-\Delta I} \, , \quad \Delta I=I+\log Z(\beta)
\ee
at the leading order in $G^{-1}_N$ expansion.

\section{{Different limits, ETH, and general proof}}

{Neither the monodromy equation \eqref{mono-eq} in the field theory nor the gravitational boundary conditions \eqref{eq1} and \eqref{eq2} can be solved exactly in terms of elementary functions. However they can be solved perturbatively in three kinds of limits, i.e., the probe limit, the heavy-shell limit and the early-time limit. In each limit, we make a comparison between the field-theory result and the gravity result and show their agreement under a Laplace transformation. Inspired by these results, we finally provide a general proof for "Field=Gravity", which is non-perturbatively correct at the leading order of the large c (small $G_N$) limit and thus go beyond all former results in the literatures  \cite{Fitzpatrick:2014vua,Fitzpatrick:2015zha,Anous:2016kss,Faulkner:2017hll}. {Along the way, we present additional analysis on ETH and unphysical solutions of the monodromy equation \eqref{mono-eq}.} 


\subsection{Probe limit}

  {Let us first consider the probe limit. The probe limit refers to the regime where $\hat e_\psi\to0$ with $e_H$ being held fixed in field theory and $m\to0$ with $\beta$ being held fixed in gravity.  }

  {On the field theory side, the probe limit is analogous to the HHLL limit for local operators with the shell operators playing similar role as the light operator. In this case, we can solve the monodromy equation as follows.} Combining \eqref{derivative} and \eqref{f-G}, we find
\be\label{K1}
\frac{\p f}{\p t_1}=K_1r_1=-\frac{3}{c}\big(-2nh_\psi+\frac{G'_{e_H}}{G_{e_H}}\big)=\hat{e}_\psi-\frac{3}{c}\frac{G'_{e_H}}{G_{e_H}}
\ee
where we use $G_{e_H}$ to {denote the correlator in the field theory} from now on, which {encodes} the information of the background state.}
Then $\rho_2$ defined in \eqref{rho12} is related to the correlator via
\be\label{rho2}
\rho_2=\frac{\sqrt{1-4(e_H-\frac{3\,G'_{e_H}}{c\, G_{e_H}})}}{2} \, ,
\ee
and the monodromy equation \eqref{mono-eq} becomes an equation for the time derivative of the  correlator.   {We may solve \eqref{mono-eq} for $G'_{e_H}/G_{e_H}$ perturbatively, considering the fact that $\hat e_\psi\ll1$. There are two types of solutions depending on the ratio of $G'_{e_H}/G_{e_H}$ to $\hat e_\psi$.} If $G'_{e_H}/G_{e_H}\sim O(\hat e_\psi)$, we would get the ordinary solution
\be\ba\label{dG:se}
\frac{G_{e_H}'}{G_{e_H}}&=-\frac{c\hat{e}_\psi\sqrt{1-4e_H}}{3}\coth\frac{\sqrt{1-4e_H}t}{2}
\\&+\frac{c\hat e_\psi^2[3+\cosh(\sqrt{1-4e_H}t)-2\sqrt{1-4e_H}t\coth\frac{\sqrt{1-4e_H}t}{2}]}{3(\cosh(\sqrt{1-4e_H}t)-1)}+O(\hat e_\psi^3)\, ,
\ea
\ee
where we have kept the first two terms in the expansion and $t=t_1-t_2$. Integrating over $t$ gives
\be\ba\label{G:se}
\log G_{e_H}=&c_s(e_H,\hat e_\psi)-\frac{2c\hat e_\psi}{3}\log\sinh\frac{\sqrt{1-4e_H}t}{2}
\\&+\frac{c\hat e_\psi^2}{3}\coth \frac{\sqrt{1-4 e_H} t }{2}\left(t \coth \frac{\sqrt{1-4 e_H} t}{2} -\frac{2}{\sqrt{1-4 e_H}}\right)+O(\hat e_\psi^3) \, ,
\ea
\ee
where $c_s$ is an integration constant. If we keep the terms up to the linearized order, the result is just the thermal two-point function in CFT$_2$ which is  expected  in the probe limit.   {This implies an important conclusion that ETH is still held by non-local shell operators  \cite{Lashkari:2016vgj,Lashkari:2017hwq} in two-dimensional holographic CFT! On the other hand, if $G'_{e_H}/G_{e_H}\sim O(1)$, there exists another family of solutions to \eqref{mono-eq}} 
\be
\ba\label{dG:se2}
\frac{G_{e_H}'}{G_{e_H}}=&\frac{c[(4e_H-1)t^2-4k^2\pi^2]}{12t^2}+\frac{16k^2\pi^2c\hat e_\psi}{3(4e_H-1)t^3-12k^2\pi^2t}
\\&-\frac{64k^2\pi^2c[3(4e_H-1)t^2+4k^2\pi^2]\hat e_\psi^2}{3[(4e_H-1)t^2-4k^2\pi^2]^3}+O(\hat e_\psi^3) \, , \quad k\in\mathbb Z \, .
\ea
\ee

It is straightforward to see that every single term (excluding the first one) in perturbation series of the solution \eqref{G:se} is singular at infinite number of points
\be
t_l=\frac{2 l \pi}{\sqrt{4e_H-1}},\quad l\in\mathbb Z,
\ee
while each term (excluding the first one) in perturbation series of the solution \eqref{dG:se2} is singular only at one point $t_k$.  These singular points $t_k$ with $k\neq0$ are the ``forbidden singularities''  \cite{Fitzpatrick:2015zha, Faulkner:2017hll} and arise as a result of the probe-limit perturbation. It is nicely argued in  \cite{Faulkner:2017hll} that for the HHLL correlator, each of these forbidden singularities would be resolved into a pair of branch-cuts when the probe corrections are taken into account. Furthermore, the vacuum block solution and additional saddle points would be sewed together to form an analytic structure of a Riemann surface with infinite sheets. For the correlator of non-local thin-shell operator in present paper,   {the picture turns out to be similar.} In particular, expanding the monodromy equation to the subleading order of $\hat e_\psi$ would lead to the same quadratic resolution to the forbidden singularity. The locations of the branch cuts appear at $t=t_k\pm i\alpha_k$ with $\alpha_k\sim O(\hat e_\psi)$.   {Despite the fact that the monodromy equation \eqref{mono-eq} for the vacuum block is exact,  its solution  is pretty complicated when $t$ is complex. Nevertheless}, the solution on real axis could be clearly shown in this case. To be more precise, solving the monodromy equation \eqref{mono-eq} for $t$ in terms of $G'_{e_H}/G_{e_H}$, we obtain a a series of {exact} expressions
\be \label{etsol}
t[k]=\frac{2\pi ki+\log\frac{3G'_{e_H}/G_{e_H}+c\hat e^2_\psi-c\hat e_\psi\sqrt{1-4e_H+\frac{12G'_{e_H}/G_{e_H}}{c}}}{3G'_{e_H}/G_{e_H}+c\hat e^2_\psi+c\hat e_\psi\sqrt{1-4e_H+\frac{12G'_{e_H}/G_{e_H}}{c}}}}{\sqrt{1-4e_H+\frac{12G'_{e_H}/G_{e_H}}{c}}}\equiv f_k(G'_{e_H}/G_{e_H}),\quad k\in\mathbb Z \, , 
\ee
When $G'_{e_H}/G_{e_H}\in(-\infty,\frac{c(4e_H-1)}{12})$, the term inside the logarithm is a pure phase\footnote{  {Note the multi-valuedness of  the logarithm function has already been taken into account in the term $2\pi ki$.} Thus the logarithm term is required to range from 0 to $2\pi$.}, thus $t\in(0,+\infty) \in \mathbb R $.   {See figure \ref{fk} for an illustration of the exact solution \eqref{etsol}.  Consistently, for a small $\hat e_\psi$, when $t$ is away from forbidden singularities, $t=f_0(G'_{e_H}/G_{e_H})$ can be well approximated by the physical solution \eqref{dG:se}, and $t=f_k(G'_{e_H}/G_{e_H})$ can be well approximated by additional solution \eqref{dG:se2} with the same $k$.  Therefore  even though the perturbative solutions \eqref{dG:se} and \eqref{dG:se2} have forbidden singularities term by term, they all can be resolved non-perturbatively by summing over full probe corrections. In particular, if we increase $t$ by staying at the real axis, the solution $G'_{e_H}/G_{e_H}$ remains smooth. The disappearance of forbidden singularities can be seen directly in the opposite heavy limit as  we are going to discuss later.}

\begin{figure}
    \centering
    \includegraphics[width=0.7\textwidth]{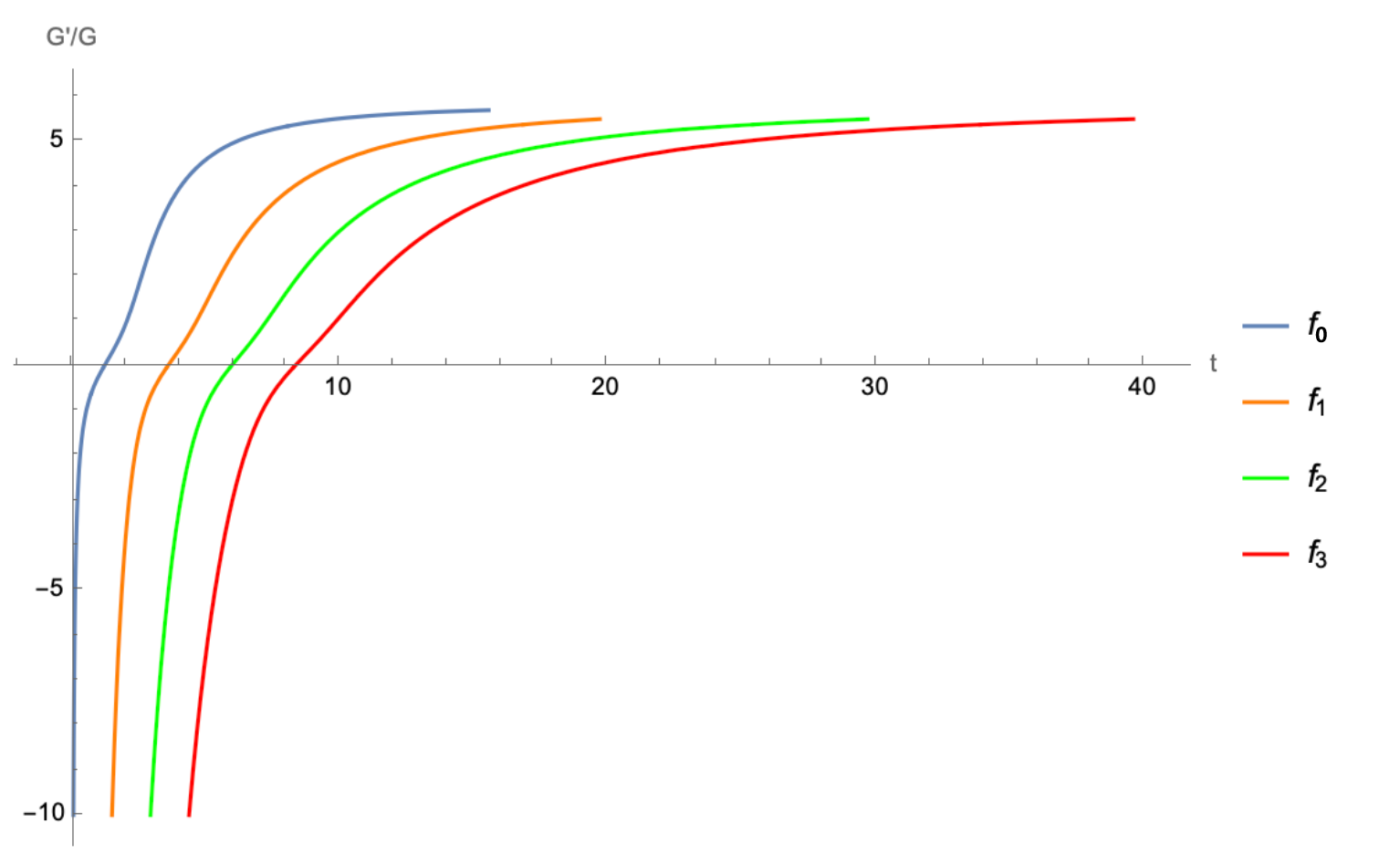}
    \caption{  Numerical plot of exact solutions for the vacuum Virasoro block of $G_{e_H}'/G_{e_H}$ \eqref{etsol} with $e_H=2,c=10,\hat e_\psi=0.1$. }
    \label{fk}
\end{figure} 
  {On the gravity side,  the probe limit corresponds to   taking $m\to0$ with $\beta$ being fixed. In this limit, the backreaction of the shell is small.} In this case, the perturbative solution is 
\be\ba
r_-&=\frac{2\pi}{\beta}+4Gm\left(\frac{1}{\pi}-\frac{t\cot\frac{\pi t}{\beta}}{\beta}\right)
\\&+\frac{4 G^2 \left(\beta ^2 \left(\pi  (2 t-\beta ) \cot \frac{\pi  t}{\beta }-2 \beta \right)-\pi ^2 t \left(\beta  (\beta -2 t)+2 \pi  t (t-\beta ) \cot \frac{\pi  t}{\beta }\right) \csc ^2\frac{\pi  t}{\beta }\right)}{\pi ^3 \beta ^2}+O(m^3)
\\ r_+&=r_-+4Gm\cot\left(\frac{\pi t}{\beta}\right)+\frac{4 G^2 m^2\left(\beta  (\beta -2 t)+2 \pi  t (t-\beta ) \cot \frac{\pi  t}{\beta }\right) \csc ^2\frac{\pi  t}{\beta }}{\pi  \beta }+O(m^3)
\ea
\ee
Substituting this into the action \eqref{action} and using \eqref{Gbeta}, we get
\be\ba\label{G:sm}
\log \left[Z(\beta)G_\beta(t)\right]\approx&\frac{\pi^2}{2G\beta}+2m\left(1+\log\frac{\pi\csc\frac{\pi t}{\beta}}{\beta}\right)
\\&+\frac{2Gm^2\left[\beta^2+\pi^2t(\beta-t)+\pi\beta(\beta-2t)\cot\frac{\pi t}{\beta}-\pi^2t(\beta-t)\csc^2\frac{\pi t}{\beta}\right]}{\pi^2\beta}+O(m^3)
\ea
\ee
Taking the derivative with respect to time $t$ gives
\be\ba\label{dG:sm}
\frac{G'_\beta(t)}{G_\beta(t)}\approx&-\frac{2\pi m\cot\frac{\pi t}{\beta}}{\beta}
\\&+\frac{2Gm^2 \left(2 \cot \frac{\pi  t}{\beta } \left(\pi ^2 t (\beta -t) \csc ^2\frac{\pi  t}{\beta }-\beta ^2\right)+\pi  \beta   (\beta -2 t) \left(1-2 \csc ^2\frac{\pi  t}{\beta }\right)\right)}{\pi  \beta ^2}+O(m^3)
\ea\ee

Note that on the gravity side with the temperature being fixed, the thermal two-point function of the thin-shell operators is computed in a canonical ensemble, while on the field side with the energy fixed, the correlation function is computed in a microcanonical ensemble.   {Thus the} correlators on two sides are supposed to be related by a Laplace transformation
\be\label{int}
Z(\beta)G_\beta(t)=\int dE_H \, \rho(E_H)e^{-\beta E_H}G_{e_H}(t) \, ,
\ee
where
\be 
E_H=\frac{c(4e_H-1)}{12} \; \longleftrightarrow \; E_H=2 h_H-\frac{c}{12}
\ee 
comes from standard AdS/CFT dictionary  \cite{Fitzpatrick:2015zha} and   {the term $-c/12$ is the Casimir energy of 2d CFT on a cylinder.}
Instead of checking whether $G_{e_H}(t)$ in \eqref{G:se} and $G_\beta(t)$ in \eqref{G:sm} satisfy \eqref{int}, we compare their derivatives \eqref{dG:se} and \eqref{dG:sm} for convenience. Performing the inverse Laplace transformation to \eqref{int}, we get
\be
\rho(E_H)G_{e_H}(t)=\int d\beta \, Z(\beta)G_\beta(t)e^{\beta E_H}.
\ee
It is difficult to perform the above integral exactly, but for the leading-order result in $G^{-1}$ expansion considering the saddle-point value is enough, so we get
\be
\rho(E_H)G_{e_H(t)}\approx Z(\beta_*)G_{\beta_*}e^{\beta_*E_H}
\ee
where the saddle point $\beta_*$ satisfies
\be\label{saddle-eq}
E_H+ \p_{\beta} \log\left[Z_\beta G_\beta(t)\right]\big|_{\beta=\beta_*}=0.
\ee
We also make the assumption that  the contribution from the high-energy states whose spectral density can be approximated by the Cardy formula in the large $c$ limit dominates  so that
\be
\rho(E_H)\approx e^{S(E_H)},\quad S(E_H)=2\pi\sqrt{\frac{c}{3}E_H}
\ee 
which is valid for $E_H>\frac{c}{12}$.  Solving \eqref{saddle-eq} gives
\be\ba\label{saddle:sm}
\beta_*=&\frac{2\pi}{\sqrt{4e_H-1}}+4Gm\left(\frac{2}{4e_H-1}-\frac{t\cot\frac{\sqrt{4e_H-1}t}{2}}{\sqrt{4e_H-1}}\right)
\\&+\frac{16G^2m^2[\sqrt{4e_H-1}t+\sin\sqrt{4e_H-1}t-(4e_H-1)t^2\cot\frac{\sqrt{4e_H-1}t}{2}]}{(4e_H-1)^{3/2}(\cos\sqrt{4e_H-1}t-1)}+O(m^3)
\ea
\ee
where   {we have used the fact that  $c=\frac{3}{2G}$.} Substituting  this saddle value into \eqref{dG:sm}, we find term by term agreement with \eqref{dG:se} provided the identification
\be\label{iden}
m=\frac{c\hat e_\psi}{3} \; \longleftrightarrow m=2 n h_{\psi}  \; .
\ee
The validity regime of the Cardy formula $E_H>\frac{c}{12}$ implies that
\be
\beta_*<2\pi+4Gm(1-t\cot\frac{t}{2})+O(m^2),
\ee
which should be understood as the regime   {where the  black hole geometry dominates the Euclidean action.} For a larger inverse temperature $\beta$, {the contribution from vacuum state  should dominate  the integral \eqref{int}. On the gravity side, this hints that}
the thermal AdS geometry would dominate over the black hole and a phase transition occurs.

\subsection{Heavy-shell limit}

The heavy-shell limit is opposite to the probe limit such that $\hat e_\psi \to\infty$ in the field theory and $m\to\infty$ in the gravity with $e_H$ and $\beta$ being fixed as well. In this limit, the backreaction of the shell would be very large.  The equations on both sides are easier to solve under this limit than {the ones in} the small $\hat e_\psi$ limit, so we work out a few more terms perturbatively to check consistency.

Under the heavy-shell limit, solving the monodromy equation \eqref{mono-eq} up to $O(\hat e_\psi^{-3})$ gives
\be\label{dG:le}
\frac{G'_{e_H}}{G_{e_H}}=\frac{c[(4e_H-1)t^2-4k^2\pi^2]}{12t^2}+\frac{4k^2\pi^2c}{3t^3\hat e_\psi}-\frac{4k^2\pi^2c}{t^4\hat e_\psi^2}+\frac{k^2\pi^2c[4(24-k^2\pi^2)+3(1-4e_H)t^2]}{9t^5\hat e_\psi^3}+O(\hat e_\psi^{-4})
\ee
with $k$ being an arbitrary integer. Here the analytic form of the correlator deviates significantly from the usual thermal one \eqref{dG:se}. More precisely, the leading term of the correlator grows in an exponential way at late time after integrating \eqref{dG:le} over $t$.

On the gravity side,  the solution is given by 
\be\ba
&r_-=\frac{2\pi}{\beta-t}-\frac{2\pi}{G(\beta-t)^2m}+\frac{2\pi}{G^2(\beta-t)^3m^2}-\frac{4\pi t^2(3-\pi^2)+3\pi^3 \beta (2t-\beta)}{6G^3t^2(\beta-t)^4m^3}+O(m^{-4}),\\
&r_+=\frac{2\pi}{t}-\frac{2\pi}{Gt^2m}+\frac{2\pi}{G^2t^3m^2}+\frac{\pi[4t^2(\pi^{2}
-3)+\beta(2t-\beta)(12-\pi^2)]}{6G^3t^4(\beta-t)^2m^3}+O(m^{-4}).
\ea
\ee
The resulting action is then
\be\ba
\log[Z(\beta)G_{\beta}]=& 2 m\log{(G m)}+ [\frac{\pi^2\beta}{2Gt(\beta-t)}-\frac{\pi^2[(\beta-t)^2+t^2]}{2G^2t^2(\beta-t)^2m}+\frac{\pi^2[(\beta-t)^3+t^3]}{2G^3t^3(\beta-t)^3m^2}
\\&+\frac{-24\pi^2[t^4+(\beta-t)^4]+\pi^4[(\beta-t)^4+t^4+6t^2(\beta-t)^2]}{48G^4t^4(\beta-t)^4m^3}+O(m^{-4}).
\ea
\ee
Taking time derivative gives
\be\ba\label{dG:lm}
\frac{G_\beta'(t)}{G_\beta(t)}=& \textcolor{cyan}{-}  \frac{\pi^2}{2G}\left(\frac{1}{t^2}-\frac{1}{(\beta-t)^2}\right)+\frac{\pi ^2 \left(\frac{1}{t^3}-\frac{1}{(\beta -t)^3}\right)}{G^2 m}+\frac{3 \pi ^2 \left(\frac{1}{(\beta-t )^4}-\frac{1}{t^4}\right)}{2 G^3 m^2}
\\&+\frac{\pi^2\left(\frac{24-\pi^2}{t^5}-\frac{24-\pi^2}{(\beta-t)^5}+\frac{3\pi^2}{t^2(\beta-t)^3}-\frac{3\pi^2}{t^3(\beta-t)^2}\right)}{12G^4m^3}+O(m^{-4}).
\ea
\ee

To compare  \eqref{dG:le} and \eqref{dG:lm}, we again solve \eqref{saddle-eq} for $\beta_*$ firstly and find
\be
\beta_*=t+\frac{2\pi}{\sqrt{4e_H-1}}-\frac{1}{Gm}+\frac{12\pi^2+(4e_H-1)t^2}{48G^3m^3t^2}+O(m^{-4}) \, .
\ee
Substituting this saddle value into \eqref{dG:lm} and the result again matches with \eqref{dG:le} in the field theory for $k=1$  under the identification \eqref{iden}.

\subsection{Early-time limit}

  {Another tractable and important limit is the early-time limit $t\to0$ where the two shells are close to each other. We discuss this limit in this subsection.} 

Similar to the case of small $\hat e_\psi$ limit, the solutions in $t\to0$ limit can be classified according to   {the ratio of $G_{e_H}'/G_{e_H}$ to $t$.} The results are summarized below.
\begin{itemize}
\item The first one is $G'_{e_H}/G_{e_H}\sim O(1)$, and in this case we get a constant solution
\be
\rho_2=0\to \frac{G_{e_H}'}{G_{e_H}}=\frac{c(4e_H-1)}{12}.
\ee
This trivial solution also appears in  previous cases by setting $k=0$ in \eqref{dG:se2} and \eqref{dG:le}.
\item The second case is that $G'_{e_H}/G_{e_H}$ is of order $O(t^{-1})$, and we solve first few orders
\be\ba\label{dG-t0}
\frac{G'_{e_H}}{G_{e_H}}&=-\frac{2c\hat e_\psi}{3t}+\frac{c\hat e_\psi^2}{9}-\frac{c[4\hat e_\psi^3+15\hat e_\psi(1-4e_H)]}{270}t+\frac{c\hat e_\psi^2[4\hat e_\psi^2+63(1-4e_H)]}{2835}t^2
\\&-\frac{c\hat e_\psi[16\hat e_\psi^4-315(1-4e_H)^2+2280\hat e_\psi^2(1-4e_H)]}{340200}t^3+O(t^4).
\ea\ee
Upon integrating over $t$, this would give the {familiar power-law behavior of correlator near the OPE singularity which in our case corresponds to the  early-time limit with } $G_{e_H}\sim t^{-\frac{2c\hat e_\psi}{3}}$.
\item There exists a third choice which is  $G'_{e_H}/G_{e_H}\sim O(t^{-2})$. Solving \eqref{mono-eq} perturbatively gives
\be
\ba
\frac{G'_{e_H}}{G_{e_H}}=&-\frac{k^2\pi^2c}{3t^2}-\frac{4c\hat e_\psi}{3t}+\frac{c(4e_H-1+\frac{16\hat e_\psi^2}{k^2\pi^2})}{12}
\\&-\frac{c\hat e_\psi[3(4e_H-1)+4\hat e_\psi^2(12-k^2\pi^2)]}{9k^4\pi^4}t+O(t^2)
\ea
\ee
with $k\in\mathbb Z$.
\end{itemize}

On the gravity side, the solutions with $t\to0$ are given by
\be
\ba
r_-&=\frac{2\pi}{\beta}+\frac{4\pi Gm}{3\beta^2}t^2-\frac{32\pi G^2m^2}{45\beta^2}t^3+\frac{4\pi Gm(21\pi^2+210G\beta m+76G^2\beta^2m^2)}{945\beta^4}t^4\\
&-\frac{64\pi G^2m^2(45\pi^2+210G\beta m+28G^2\beta^2m^2)}{14175\beta^4}t^5+O(t^6)\\
r_+&=4\sqrt{\frac{Gm}{t}}+\frac{3\pi^2-4G^2\beta^2m^2}{6\sqrt{Gm}\beta^2}\sqrt{t}-\frac{\sqrt{Gm}(45\pi^4+840\pi^2G^2\beta^2m^2-176G^4\beta^4m^4)}{1440G^2\beta^{4}m^2}t^{3/2}
\\&+\frac{\sqrt{Gm}[315\pi^6+5460\pi^4G^2\beta^2m^2-1088G^6\beta^6m^6+336\pi^2G^3\beta^3m^3(160+101G\beta m)]}{80640G^3\beta^6m^3}t^{5/2}\\
&-\frac{\sqrt{Gm}}{38707200G^4\beta^8m^4}t^{7/2}\times\left[[23625\pi^8+478800\pi^6G^2\beta^2m^2+71936G^8\beta^8m^8\right.
\\& \left.+10080\pi^4G^3\beta^3m^3(320+489G\beta m)+3840\pi^2G^5\beta^5m^5(11424+2339G\beta m)] \right]
\ea
\ee
Then we find
\be\ba\label{G:t0}
\log[Z(\beta)G_\beta(t)]=&\frac{\pi^2+4G\beta m (1-\log t)}{2G\beta}+\frac{2Gm^2t}{3}+\frac{(15\pi^2m-4G^2\beta^2m^3)t^2}{45\beta^2}\\
&-\frac{8(63\pi^2Gm^2-4G^3\beta^2m^3)}{2835 \beta^2}t^3 \\&+\frac{[315\pi^4-16G^4\beta^4m^4+60\pi^2G\beta m(105+38G\beta m)]mt^4}{28350\beta^4}
+O(t^5)
\ea
\ee
Taking time derivative gives
\be\ba\label{dG:t0}
\frac{G'_{\beta}(t)}{G_{\beta}(t)}&=-\frac{2m}{t}+\frac{2Gm^2}{3}+\left(\frac{2m\pi^2}{3\beta^2}-\frac{8G^2m^3}{45}\right)t
\\&+\frac{8Gm^2}{945}\left(4G^2m^2-\frac{63\pi^2}{\beta^2}\right)t^2\\
&+\frac{2m[315\pi^{4}-16G^4\beta^4m^{4}+60\pi^2G\beta m(105+38G\beta m)]}{14175\beta^4}t^3+O(t^4)
\ea
\ee

Solving \eqref{saddle-eq} for $\beta_*$ with \eqref{G:t0}, we get
\be
\beta_*=\frac{2\pi}{\sqrt{4e_H-1}}+\frac{2Gmt^2}{3}-\frac{16G^2m^2t^3}{45}+\frac{Gm[21(4e_H-1)+304G^2m^2]t^4}{1890}+O(t^5)
\ee
Substituting above solution into \eqref{dG:t0}, the result matches with \eqref{dG-t0} perfectly under the identification \eqref{iden}.

To conclude, we analyze the perturbative solutions to the monodromy equation \eqref{mono-eq} in the field theory and the on-shell equation \eqref{eq1},\eqref{eq2} in the gravity in three different limits. The resulting time derivatives of the correlators on both sides agree with each other under an inverse Laplace transformation to the leading order in $G^{-1}$ expansion. 

\subsection{Remarks On ETH Ansatz}
The thin-shell operator is studied not only for its convenience in constructing dual semiclassical geometry, but also it comforts to the form of the ETH ansatz.   {The ETH ansatz} is an assumption on the matrix element of the operator
\be
\mathcal V_{nm}:=\langle E_n|\mathcal V|E_m\rangle=e^{-f(E_n,E_m)/2}R_{nm}
\ee
where   {the envelop function $f(E_n,E_m)$ is a symmetric smooth function of $E_n,E_m$} and $R_{nm}$ is a complex Gaussian random variable.
With this ansatz, the thermal two-point function  $G_\beta(t)$ can be written as the double Laplace transformation of $e^{-f(E_1,E_2)}$
\be
\ba\label{ETH}
G_\beta(t)
=\frac{1}{Z(\beta)}\int dE_1dE_2e^{S(E_1)+S(E_2)-(\beta-t)E_1-tE_2-f(E_1,E_2)}.
\ea
\ee
To the leading order in $G^{-1}$ expansion, this can be further approximated by its saddle-point value so that
\be\ba\label{ETH-saddle}
G_\beta(t)&\approx\frac{1}{Z(\beta)}e^{S(E^*_1)+S(E^*_2)-(\beta-t)E_1^*-tE_2^*-f(E_1^*,E_2^*)}
\\&=\frac{1}{Z(\beta)}e^{-(\beta-t)F(\beta_{E_1^*})-tF(\beta_{E_2^*})+\frac{\p_{E_1^*}f}{\beta_{E_1^*}}S(E_1^*)+\frac{\p_{E_2^*}f}{\beta_{E_2^*}}S(E_2^*)-f(E_1^*,E_2^*)}
\ea
\ee
where in the second line we have used the thermodynamic relation $F(\beta^*_{E_i}):=E_i^*-\beta^{-1}_{E_i^*}S(E_i^*),\beta_{E_i^*}:=S'(E_i^*)$ for $i=1,2$ and the saddle-point equation for $E_1^*,E_2^*$
\be
\p_{E_1^*}f(E_1^*,E_2^*)=\beta_{E_1^*}-\beta+t,\quad \p_{E_2^*}f(E_1^*,E_2^*)=\beta_{E_2^*}-t.
\ee
Comparing \eqref{ETH-saddle} with \eqref{Gbeta} and \eqref{act3}, we find that 
\be
I_s=\frac{\p_{E_1^*}f}{\beta_{E_1^*}}S(E_1^*)+\frac{\p_{E_2^*}f}{\beta_{E_2^*}}S(E_2^*)-f(E_1^*,E_2^*)
\ee
provided that we identify $E_1^*=M_-,E_2^*=M_+$.  Moreover, we note that in \eqref{ETH-saddle}, $G_\beta(t)$ depends on time both explicitly and implicitly through $E_i^*$. Since $E_i^*$ satisfies the saddle-point equation, the time derivative of $G_\beta(t)$ can be taken directly.   {Using} the first line of \eqref{ETH-saddle}, we have
\be\label{time-Gb}
\frac{G'_\beta(t)}{G_\beta(t)}=E_1^*-E_2^*=M_--M_+.
\ee
The above relation between the time derivative of thermal correlation and the mass difference of two black holes can indeed be checked directly using the perturbative solutions we obtained before. 

For the microcanonical correlation function $G_{e_H}(t)$, the ansatz \eqref{ETH} can be applied similarly.
 Combining \eqref{Gbeta} and \eqref{ETH}, we would have
\be\label{Geh}
G_{e_H}(t)=\int dEe^{S(E)+t(E_H-E)-f(E_H,E)}.
\ee
Again, at the semiclassical level, the saddle-point approximation gives
\be\label{Geh-saddle}
G_{e_H}(t)\approx e^{S(E^*)+t(E_H-E^*)-f(E_H,E^*)}
\ee
with $E^*=E^*(E_H,t)$ satisfying
\be
\beta_{E^*}=t-\p_{E^*}f(E_H,E^*).
\ee
Taking the time derivative of \eqref{Geh-saddle}, we would get
\be\label{time-Geh}
\frac{G'_{e_H}}{G_{e_H}}=E_H-E^*.
\ee
Comparing \eqref{rho2} with \eqref{time-Geh}, we find that the time derivative of the correlaor computed using the monodromy method fits into the form suggested by the ETH ansatz provided we identify\footnote{Note that $E_H=\frac{c(4e_H-1)}{12}$.}
\be\label{E-rho}
E^*=-\frac{c\rho_2^2}{3}.
\ee
In this sense, we can interpret $\rho_2$   {as the label of} the energy of the state which contributes most to the microcanonical correlator in the integral \eqref{Geh}. 

Similarly, if we perform a Laplace transformation from $G_{e_H}(t)$ to $G_\beta(t)$, the saddle-point equation can be solved to give $e_H^*=e_H^*(\beta,t)$. Then the identification between the time derivative of $G_\beta(t)$ and $G_{e_H^*}(t)$ implies
\be\label{M-E}
M_-=E_H,\quad M_+=E^*
\ee
where the right hand side is evaluated at $e_H=e_H^*$. Note that \eqref{E-rho} is obtained by combining ETH ansatz \eqref{ETH} and the monodromy method, while \eqref{M-E}  is obtained by combining ETH ansatz with the gravity calculation. 
From our previous calculation, it can be checked explicitly that if we substitute the saddle value $\beta=\beta_*$ into the solution of $r_-$ and $r_+$, we would have
\be\label{relation}
M_-\big|_{\beta_*}=E_H,\quad M_+\big|_{\beta_*}=-\frac{\rho_2^2}{2G},
\ee
which is nothing but \eqref{M-E} using \eqref{E-rho}. This suggests that our perturbative solutions from the field theory and the gravity would give consistent ETH ansatz.  

\subsection{General Proof}\label{proof}

In this subsection, we provide a general proof that up to an inverse Laplace transformation,   {the contribution of the vacuum Virasoro block to} microcanonical correlator obtained by the generalized monodromy method can indeed be ``matched'' with the thermal one computed in gravity non-perturbatively. Moreover, the relations \eqref{relation} can also be consistently verified.

The explicit expression for the  on-shell gravity action \eqref{action} is given by
\be\ba\label{grav-G}
&\log [Z(\beta)G_{\beta}]=-I=\frac{r_+^2t+r_-^2(\beta-t)-16Gm\log 2+16Gm\log\sqrt{r_+^2+(\frac{r_+^2-r_-^2}{8Gm}-2Gm)^2}}{8G}.
\ea\ee
  {Taking time derivative of \eqref{grav-G} and keeping in mind that $r_\pm$ are the functions of $(\beta,t)$ after solving \eqref{grav-eq}, we get}
\be\label{Gbeta:dt}
\frac{G'_\beta}{G_\beta}=\frac{r_+^2-r_-^2+(r_+^2)'t+(r_-^2)'(\beta-t)+\frac{16Gm[(r_-^2)'(16G^2m^2+r_-^2-r_+^2)+(r_+^2)'(16G^2m^2-r_-^2+r_+^2)]}{pq}}{8G}
\ee
where $r'_\pm=\p_t r_\pm$ and
\be
p=16G^2m^2+(r_--r_+)^2,\quad q=16G^2m^2+(r_-+r_+)^2.
\ee
We then take the time derivative of \eqref{grav-eq} and solve the algebraic equations for $r'_\pm$, 
\be
\ba\label{rpm:dt}
r_-'=&\frac{r_-[pqt+16Gm(16G^2m^2+r_-^2-r_+^2)]}{pq(\beta-t)t+8Gm[32Gm+\beta(p+q)]},\\
r_+'=&-\frac{r_+[pq(\beta-t)+16Gm(16G^2m^2-r_-^2+r_+^2)]}{pq(\beta-t)t+8Gm[32Gm+\beta(p+q)]}.
\ea
\ee
Substituting \eqref{rpm:dt} into \eqref{Gbeta:dt} gives
\be\label{relation2}
r_-^2-r_+^2-\frac{8GG'_\beta}{G_\beta}=0.
\ee

We may compute the microcanonical correlator $\tilde G_{e_H}(t)$ in gravity  via the inverse Laplace transformation of $Z(\beta)G_\beta(t)$. With the saddle-point approximation, it is given by
\be\label{tilde-G}
\log\tilde G_{e_H}(t)\approx \log [Z(\beta_*)G_{\beta_*}]+\frac{\beta_*c(4e_H-1)}{12}-\frac{\pi c\sqrt{4e_H-1}}{3}
\ee
where $\beta_*(e_H,t)$ satisfies
\be\label{saddle-beta}
\frac{c(4e_H-1)}{12}+\frac{\p_\beta [Z(\beta)G_\beta(t)]}{Z(\beta)G_\beta(t)}\big|_{\beta_*}=0 \, ,
\ee
which is exactly the same condition as \eqref{saddle-eq}. {Here we would like to compute the correlator exactly}, not perturbatively. {According to the definition of $\tilde G_{e_H}(t)$ in \eqref{tilde-G}, one can see straightforwardly that the microcanonical correlator $\tilde G_{e_H}$ and the canonical correlator $G_{\beta_*}$ in gravity have the same time derivative 
\be \label{xin1}
\frac{\tilde G'_{e_H}}{\tilde G_{e_H}}=\frac{G'_{\beta_*}}{G_{\beta_*}} \, .
\ee Thus we can use $G'_{\beta_*}/G_{\beta_*}$ rather than the microcanonical correlator in gravity to complete the proof.  As shown above,}   {the derivative of $Z(\beta)G_{\beta}(t)$ with respect to $\beta$ }can also be obtained from \eqref{grav-eq}, i.e., solving $\p_\beta r_\pm$ by taking the derivative of \eqref{grav-eq} with respect to $\beta$. Plugging the results into \eqref{saddle-beta}, we find that $\beta_*$ satisfies
\be  \label{relation3}
r_-^{2}(\beta_*,t)=4e_H-1=-4\rho_1^2.
\ee

Now we are ready to prove that the microcanonical correlator $G'_{e_H}/G_{e_H}$ in the field theory equals to the correlator $G'_{\beta_*}/G_{\beta_*}$ in gravity, i.e.,
\be  \left.\frac{G'_{e_H}}{G_{e_H}}\right|_{\mbox{\tiny Field}}=\left.\frac{G'_{\beta_*}}{G_{\beta_*}}\right|_{\mbox{\tiny Gravity}}. \label{fldgrveq} \ee
We demonstrate the  equality by showing that the monodromy equation \eqref{mono-eq} and the gravity condition \eqref{grav-eq}\footnote{Only the second equation in \eqref{grav-eq} is { relevant}, because $r_-$ has already been fixed by the condition \eqref{relation3}.} are exactly the same.  Using \eqref{relation2}, \eqref{relation3} and \eqref{xin1}, we can get
the following relation\be \label{relation4} r_+^{2}(\beta_*,t)=-4\tilde\rho_2^2 \,. \ee
where 
\be\label{tilderho2}
\tilde\rho_2\equiv\frac{\sqrt{1-4(e_H-\frac{3\tilde G_{e_H}'}{c\tilde G_{e_H}})}}{2}.\ee
Using the second equation of gravity condition \eqref{grav-eq} with  $t=t_1-t_2$, we have 
\begin{align}  \label{deng2}
    e^{2\tilde\rho_2(t_1-t_2)}&=e^{ i r_+ (t_1-t_2)} = 1-2 \sin^2{\frac{r_+(t_1-t_2)}{2}} + 2 i \sin{\frac{r_+(t_1-t_2)}{2}} \cos{\frac{r_+(t_1-t_2)}{2}} \nonumber \\
    &=\frac{16 G^2 m^2+r_{-}^2-r_{+}^2-8 i G m r_{+}  }{16 G^2 m^2+r_{-}^2-r_{+}^2+8 i G m r_{+} }\bigg\vert_{\beta=\beta_{*}}.
\end{align} 
Furthermore,  with the identification $\hat e_\psi=3m/c=2Gm$, the above equation can be written as
\be \label{new2}
e^{2\tilde\rho_2(t_1-t_2)}= \frac{4 \hat{e}_\psi^2+r_{-}^2-r_{+}^2-4 i \hat{e}_\psi r_{+}}{4 \hat{e}_\psi^2+r_{-}^2-r_{+}^2+4 i \hat{e}_\psi r_{+}}\bigg\vert_{\beta=\beta_{*}}.
\ee
{ Using the relations \eqref{relation3}, \eqref{relation4} and \eqref{tilderho2}, \eqref{new2} becomes a non-linear equation of $\tilde G'_{e_H}/\tilde G_{e_H}$. It is straightforward to show that \eqref{new2} takes the same form as \eqref{mono-eq} which means that
$\tilde G'_{e_H}/\tilde G_{e_H}$ satisfies the same monodromy equation as $G'_{e_H}/G_{e_H}$. Therefore, with the initial condition at the OPE singularity $t\to0$ fixed, we can conclude that 
\be \label{xin4}
\frac{G'_{e_H}}{G_{e_H}}=\frac{\tilde G'_{e_H}}{\tilde G_{e_H}}=\frac{G'_{\beta_*}}{G_{\beta_*}},
\ee 
where we have used \eqref{xin1}, and we complete the proof.   
As a side remark, we note that \eqref{xin4} implies
\be \label{new3} r_+^{2}(\beta_*,t)=-4\rho_2^2 \, . \ee
This equation, together with \eqref{relation3}, are exactly the same relations in \eqref{relation} that we get in general analysis of the ETH anstaz.  }


\subsection{Singular Solutions}
In previous discussions, we have obtained {an infinite number of solutions to the monodromy equation. Only  one of them is physical, and leads to the result in agreement with the result in gravity.} In this section, we focus on those additional solutions and give them a geometric dual.
On the field theory side, recall that the mondromy matrix $M$ should obey the trace condition
\be
\text{Tr}M=-2\cos(\pi \Lambda_h),\quad h=\frac{c}{24}(1-\Lambda_h^2)
\ee
with $h$ being   {the scaling dimension of the operators in the propagating channel. For the identity operator} $h=0$, this becomes $\text{Tr}M=2$. However, there are infinitely many $h'\neq0$ that satisfies the same monodromy condition as   {the operator $h$ }, which are determined by
\be
\Lambda_h=\Lambda_{h'_k}+2k,\quad k\in\mathbb Z.
\ee
For $h=0$, we  have $h'_k=-\frac{c}{6}k(k+1)$ and  the monodromies of these solutions  wind around $k$ times before going back to trivial. We expect that the conformal blocks   {corresponding to these operators which have negative dimensions  produce the additional solutions to the monodromy equation. } 

 In the following,   {we propose gravitational configurations  whose on-shell actions reproduce  those additional solutions in the field theory.} Instead of requiring the periodicity $t_\pm\sim t_\pm+\beta_\pm$ to satisfy \eqref{eq1} and \eqref{eq2}, we impose the following relations
\be
\ba\label{grav-eq-k}
&\beta_-=\beta-t+\Delta t_-,\quad t-\Delta t_+=k\beta_+,\quad &\text{for}~M_--M_+>2Gm^2,
\\&\beta_-=\beta-t+\Delta t_-,\quad t+\Delta t_+=(k+1)\beta_+,\quad &\text{for}~M_--M_+<2Gm^2.
\ea
\ee
The physical interpretation of above relations is that the dust particles travel around the right black hole $k$ times before they end on the boundary.  We can use the same method as {the one in} section \ref{proof} to show that the on-shell actions evaluated on the solutions to \eqref{grav-eq-k} also solves the monodromy equation after the inverse Laplace transformation. In the following, we just take $m\to0$ as a simple example to show that
the additional solutions \eqref{dG:se2} to the monodromy can be reproduced in this way. Solving \eqref{grav-eq-k} perturbatively in small $m$ gives rise to 
\be\ba
r_-=&\frac{2\pi}{\beta-t}+\frac{8Gt^2}{\pi(k\beta-(k+1)t)(k\beta-(k-1)t)}m
\\&+\frac{32G^2t^3(\beta-t)[2k^2\beta^3-5k^2\beta^2t+4k^2\beta t^2-(k^2-1)t^3]}{\pi^3(k\beta-(k+1)t)^3(k\beta-(k-1)t)^3}m^2+O(m^3),\\
r_+=&\frac{2k\pi}{t}+\frac{8Gk(\beta-t)^2}{\pi(k\beta-(k+1)t)(k\beta-(k-1)t)}m
\\&+\frac{32G^2kt(\beta-t)^3[k^2\beta^3-3k^2\beta^2t+(3k^2+1)\beta t^2-(k^2-1)t^3]}{\pi^3(k\beta-(k+1)t)^3(k\beta-(k-1)t)^3}m^2+O(m^3).
\ea\ee
Substituting this into the action \eqref{action}, we find
\be\ba
\log[Z(\beta)G_\beta(t)]\approx&\frac{\pi^2(k^2\beta-(k^2-1)t)}{2Gt(\beta-t)}+2\left(\log\frac{\pi^2\left(\frac{k^2}{t^2}-\frac{1}{(\beta-t)^2}\right)}{4Gm}+2\right)m\\
&+\frac{8Gt(\beta-t)[k^2\beta^3-3k^2\beta^2t+3k^2\beta t^2-(k^2-1)t^3]}{\pi^2(k\beta-(k+1)t)^2(k\beta-(k-1)t)^2}m^2+O(m^3)
\ea
\ee
with its time derivative given by
\be\ba\label{dG:sm:k}
\frac{G'_\beta(t)}{G_\beta(t)}\approx&\frac{\pi\left(\frac{1}{(\beta-t)^2}-\frac{k^2}{t^2}\right)}{2G}-\frac{4m\left(\frac{k^2}{t^3}+\frac{1}{(\beta-t)^3}\right)}{\frac{k^2}{t^2}-\frac{1}{(\beta-t)^2}}
\\&+\frac{8Gm^2\left(\frac{k^4}{t^6}+\frac{1}{(\beta-t)^6}+\frac{k^2(3\beta^2+2\beta t-2t^2)}{t^4(\beta-t)^4}\right)}{\pi^2\left(\frac{k^2}{t^2}-\frac{1}{(\beta-t)^2}\right)^3}+O(m^3).
\ea
\ee
Solving the saddle-point equation \eqref{saddle-eq} for the inverse Laplace transformation gives
\be
\beta_*=t+\frac{2\pi}{\sqrt{4e_H-1}}+\frac{16Gt^2m}{(4e_H-1)t^2-4k^2\pi^2}+\frac{2048G^2k^2\pi^2t^3m^2}{[(4e_H-1)t^2-4k^2\pi^2]^3}+O(m^3).
\ee
Substituting it into \eqref{dG:sm:k}, we get the same result as \eqref{dG:se2}. 

\section{Conclusion}
In this paper, we studied the correlation function of thin-shell operators using the monodromy method. In particular, we considered the contributions from the vacuum conformal block. Requiring the monodromy matrix to be trivial gives us an exact algebraic equation for the time derivative of the correlator. In the whole procedure, we did not make any approximation such as   {the probe limit usually used in dealing with correlation} function of four local operators.   { This} equation can not only be solved perturbatively in the probe limit, but also can be easily solved in the opposite limit where the shell operators are much heavier than the local operators that create the heavy state.  It is worth noting that even in the probe limit, we still require that $\hat e_\psi\gg c^{-1}$ so that our result  remains at the semiclassical order.

Motivated by the simplicity of the monodromy equation,  we studied its solutions in three meaningful limits: the probe limit, the heavy-shell limit, and the early time limit. In each limit, we found one physical solution as well as an infinite class of additional solutions   {which are unphysical. These additional saddles are expected to originate from the ambiguity in the monodromy condition, since there exists an infinite class of intermediate operators with negative dimensions which share the same monodromy as the identity operator.}  In the probe limit, the leading order of physical solution behaves like the usual thermal two-point function as expected while the other solutions behave in a different way and exhibits an exponential behavior upon integrating over time. Similar results occur in the early-time limit where the physical solution indicates the power-law behavior of the correlator while the other solutions indicate an exponential-like behavior. For the heavy limit, it is a bit tricky to distinguish the physical solution from the unphysical ones and they all suggest an exponential behavior of the correlators in time. 
In the probe limit, we also saw the appearance of forbidden singularities. For the physical solution, there are infinite forbidden singularities at each order of the solutions, while there is only one forbidden singularity associated to each additional solution. We showed  explicitly that after summing over all probe corrections, these singularities do not occur and the solution is smooth on the real axis. If we consider the behavior of $G'_{e_H}/G_{e_H}$ on the whole complex plane of $t$, we expect to see the similar structure as discussed in  \cite{Faulkner:2017hll}.

Holographically, the thin-shell operator is described by a perfect fluid propagating into the bulk and backreacts on the geometry. At the semiclassical level, the thermal correlator of two shell operators is given by evaluating the Euclidean on-shell action of the backreacted   {spacetime geometry. The  backreacted spacetime} is given by two BTZ black holes gluing across the world volume  of the shell via the junction conditions, as has been addressed in  \cite{Sasieta:2022ksu}. The masses of the two black holes can be determined by properly {identifying} the thermal circle with the boundary insertion time and the time elapsed by the shell. Again, we solved the backreacted geometry and obtained the on-shell action in the three limits as what we have done in the field theory.   {Then we checked explicitly that by  using the inverse Laplace transformation the time derivative of the thermal correlation agrees exactly with the one determined by solving the monodromy equation in the field theory.}
Inspired by this observation, we provided a general proof at the level of equations to show that this agreement holds to all orders. Furthermore, we also gave a geometric interpretation for the additional unphysical solutions in solving the monodromy equation. 

There are several future directions {worth being explored further} based on our study:
\paragraph{Higher-point functions}
Although we only considered the two-point function in this paper, we expect  that the relations \eqref{int} can be proved for the correlators of $2n$ shell operators. In this case, we would have $n$ loops on which to impose  the monodromy conditions and therefore get $n$ monodromy equations. The saddle-point equation to the Laplace transformation \eqref{int} will also give one more equation.
On the gravity side, there are $n$ trajectories of dust particles.  Since there is no stable saddle point to the geometry in which the two shells cross, the $n$ trajectories will divide the spacetime into $n+1$ regions with each  being locally a BTZ {geometry}
and lead to $n+1$ equations of the identification among the boundary insertion time, the temperature of each BTZ black hole and the time elapsed by the shell.  It would be interesting to show   {if  the two sets of equations  in the field theory and gravity can be }  consistent with each other.
\paragraph{Adding angular momentum}
It is natural to generalize our discussion to the rotating BTZ black hole with angular momentum $J$. In the field theory, this requires us to consider $e_H\neq \bar e_H$ and check whether
\be
G_{\beta,J}(t)=\int de_Hd\bar e_He^{-\beta(E_H+\bar E_H)-J(E_H-\bar E_H)}G_{e_H,\bar e_H}(t)
\ee
holds or not. It would be good if one can even make further generalization by considering spinning shells so that $h_\psi\neq \bar h_\psi$.

\paragraph{Analytic continuation}
Another important problem would be the behavior of the microcanonical correlator at late Lorentzian time. In the case of local operators, the forbidden singularities in the Euclidean regime imply an exponential decay in Lorentzian regime which leads to the information loss problem \cite{Fitzpatrick:2016ive}. The resolution to the forbidden singularities or the exponential decay for the Virasoro block has been studied both analytically \cite{Fitzpatrick:2016mjq} and numerically \cite{Chen:2017yze} which showed that the Lorentzian correlator exhibits a universal power-law decay at late Lorentzian time. In our work, we have already seen the difference between the correlator of shell  operators and that of local operators in Euclidean regime. We wish to explore the behavior of the correlator in the Lorentzian regime by solving the analytic continuation of the monodromy equation in the future.

\section*{Acknowledgments}
{We thank Peng Cheng, Huajia Wang and Jieqiang Wu for helpful discussions.}
    This work is partially supported by NSFC Grant  No. 11735001, 12275004.
   
\bibliographystyle{JHEP}
\bibliography{biblio}
\end{document}